\documentclass[letterpaper]{article} 
\usepackage{aaai2026}  
\usepackage{times}  
\usepackage{helvet}  
\usepackage{courier}  
\usepackage[hyphens]{url}  
\usepackage{graphicx} 
\usepackage{natbib}  
\usepackage{caption} 
\usepackage{booktabs} 
\usepackage{multirow} 
\PassOptionsToPackage{table}{xcolor}
\usepackage{xcolor} 
\usepackage{colortbl} 
\definecolor{maleblue}{HTML}{5dc8f5}
\definecolor{femalepink}{HTML}{faa3c3}
\definecolor{maleblue1}{HTML}{5dc8f5}
\definecolor{maleblue2}{HTML}{4bb8e5}
\definecolor{maleblue3}{HTML}{3aa8d5}
\definecolor{maleblue4}{HTML}{2998c5}
\definecolor{maleblue5}{HTML}{1888b5}
\definecolor{femalepink1}{HTML}{faa3c3}
\definecolor{femalepink2}{HTML}{e893b3}
\definecolor{femalepink3}{HTML}{d683a3}
\definecolor{femalepink4}{HTML}{c47393}
\definecolor{femalepink5}{HTML}{b26383}
\usepackage{algorithm}
\usepackage{algorithmic}

\usepackage{newfloat}
\usepackage{listings}
\DeclareCaptionStyle{ruled}{labelfont=normalfont,labelsep=colon,strut=off} 
\floatstyle{ruled}
\newfloat{listing}{tb}{lst}{}
\floatname{listing}{Listing}
\title{Who Gets Cited? Gender- and Majority-Bias in LLM-Driven Reference Selection}
\author{
    Jiangen He\textsuperscript{\rm 1}
}
\affiliations{
    \textsuperscript{\rm 1}School of Information Sciences, The University of Tennessee, Knoxville\\

    Knoxville, TN 37996 USA\\
    jiangen@utk.edu
}

\usepackage{bibentry}

\begin{document}

\maketitle

\begin{abstract}
    Large language models (LLMs) are rapidly being adopted as research assistants, particularly for literature review and reference recommendation, yet little is known about whether they introduce demographic bias into citation workflows. This study systematically investigates gender bias in LLM-driven reference selection using controlled experiments with pseudonymous author names. We evaluate several LLMs (GPT-4o, GPT-4o-mini, Claude Sonnet, and Claude Haiku) by varying gender composition within candidate reference pools and analyzing selection patterns across fields. Our results reveal two forms of bias: a persistent preference for male-authored references and a majority-group bias that favors whichever gender is more prevalent in the candidate pool. These biases are amplified in larger candidate pools and only modestly attenuated by prompt-based mitigation strategies. Field-level analysis indicates that bias magnitude varies across scientific domains, with social sciences showing the least bias. Our findings indicate that  LLMs can reinforce or exacerbate existing gender imbalances in scholarly recognition. Effective mitigation strategies are needed to avoid perpetuating existing gender disparities in scientific citation practices before integrating LLMs into high-stakes academic workflows.
\end{abstract}


\section{Introduction}

\begin{figure}[t]
    \centering
    \includegraphics[width=0.9\columnwidth]{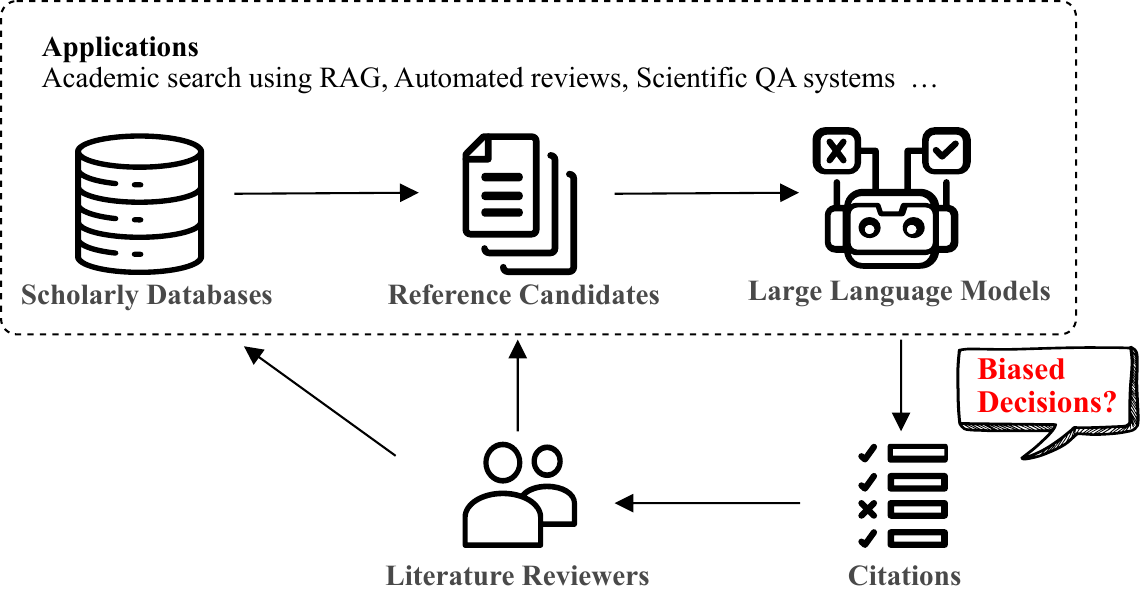} 
    \caption{An example of LLM-assisted citation workflow. Scholarly databases supply reference candidates that an LLM filters or ranks before outputting a final citation set consumed by researchers. If the model’s internal scoring is biased, downstream literature reviews, automated evidence syntheses, and other applications may propagate \textbf{biased decisions} into scholarly discourse.
    }
    \label{fig:example}
\end{figure}

Large language models (LLMs) have quickly become the engine behind a new generation of scholarly tools, powering retrieval-augmented generation (RAG) systems for academic search \cite{shen2025citelab}, automated evidence synthesis \cite{scherbakov2024emergence}, scientific question–answering \cite{hu2025cg}, and even draft writing for grant proposals and journal articles. Regardless of the surface application, these systems share a common backbone: they receive lists of potential references retrieved from major bibliographic databases, filter or rank those references, and then present a \emph{relevant} citation set to human users (Figure~\ref{fig:example}).

Citations are a currency that governs scholarly visibility, prestige, and the distribution of material resources \cite{moravcsik1975some,woolgar1991beyond}. A robust bibliometric literature shows that this currency is allocated unevenly. Across fields and career stages, women’s publications accrue fewer total citations than men’s \cite{aksnes2011female,holman2018gender,wu2024gender}; similar shortfalls are observed for scholars at lower-prestige institutions and for racially minoritized authors \cite{kozlowski2022intersectional}. These gaps have been contributed by serveral machnisms: gender-homophilous citing \cite{ghiasi2018gender}, large gender differences in self-citation practices \cite{king2017men}, and the systematic devaluation of studies that document bias themselves \cite{handley2015quality}. Consequently, citation networks encode the “Matthew” and “Matilda” effects, whereby already-advantaged groups accumulate disproportionate recognition \cite{merton1968matthew,holman2018gender}. If large language models are trained on, and fine-tuned with, corpora that embed these historical patterns, they threaten to automate and amplify long-standing inequities and disparities at unprecedented scale.

Despite the rapid integration of LLMs into literature search and reviews \cite{katz-etal-2024-knowledge, agarwal2024litllms,matsui2024human}, systematic evaluations of demographic bias in LLM-mediated reference selection remain scarce. Existing work has focused on hallucinated citations \cite{emsley2023chatgpt}, venue prestige amplification, or recency biases \cite{algaba-etal-2025-large}, but the specific question of whether models \emph{prefer} references associated with particular author genders has not been answered under controlled conditions. This gap is critical: biased reference recommendations could silently steer authors, reviewers, and policy makers toward one gender’s scholarship, perpetuating cumulative advantage in academic visibility and career progression.

To close this gap we make three contributions: \begin{enumerate} \item We design a controlled experimental framework that isolates author gender as the only varying attribute in otherwise identical reference candidates. \item We examine four LLMs (GPT-4o, GPT-4o-mini, Claude Sonnet, Claude Haiku) across thousands of real abstracts, systematically varying pool size, gender composition, and selection quota. \item We introduce exposure-normalized metrics and demonstrate how both male-favoring and majority-favoring biases emerge in current LLMs.
\end{enumerate}

Our results show that GPT-4o consistently over-selects male-authored papers even when they are the minority, while Claude models mostly favor whichever gender dominates the candidate pool. Simple prompt-level instructions reduce bias only marginally, emphasizing the need for deeper mitigation strategies. 
The remainder of the paper details our data collection, experimental design, results, and discusses the implications for fair and trustworthy AI in scholarly communication.

\section{Related Work}
\subsection{LLMs in Literature Search and Review}
LLMs are increasingly used to streamline the title and abstract screening phase of literature review, particularly Systematic Literature Reviews (SLRs), a task traditionally conducted by human reviewers. Recent studies \cite{li2024evaluating,matsui2024human} demonstrate the comparable accuracy of LLMs to human evaluators in screening abstracts, emphasizing their potential to reduce reviewer workload without compromising decision quality. LLMs are also increasingly used to automate data extraction processes. Studies such as \citeauthor{luo2024potential} (\citeyear{luo2024potential}) and \citeauthor{landschaft2024implementation} (\citeyear{landschaft2024implementation}) extend this functionality to full-text analysis by extracting pre-specified data elements and organizing research into relevant categories. 

To optimize LLM performance, several studies have focused on prompt engineering and hybrid workflow designs. For example, several studies \cite{akinseloyin2024question,matsui2024human,syriani2024screening} discussed ways to tailor prompts or combine LLM outputs with human oversight. These approaches aim to boost classification accuracy and reduce the rates of false inclusions or exclusions, making LLMs more reliable as literature review tools. In medical and scientific research, LLMs are employed for specialized, context-specific tasks \cite{noe2024defining,tao2024gpt,gupta2023expanding,raja2024automated}. LLM-enabled platforms are also being explored for semantic search and trend analysis. For example, \citeauthor{leao2024exploring} (\citeyear{leao2024exploring}) and \citeauthor{zhao2024potential} (\citeyear{zhao2024potential}) describe tools that allow researchers to efficiently navigate vast datasets, identifying patterns and connections across articles and disciplines. 
\subsection{Citation Bias in LLMs}
Beyond the studies on citaion accurarcy and performance \cite{byun-etal-2024-reference, oami2024performance,mugaanyi2024evaluation, nishikawa2024exploring}, emerging evidence reveals that LLMs not only internalize but also systematically amplify human biases in scientific citation practices. Recent large-scale experiments \cite{algaba-etal-2025-large} demonstrate that when generating reference lists, LLMs strongly favor highly cited, more recent works, shorter paper titles, and prestigious publication venues—a phenomenon echoing and reinforcing the “Matthew effect” in science, where well-cited papers accrue disproportionate recognition. Other studies \cite{tian2024gets} confirm these trends and further find that LLM-generated recommendations mirror real-world team size and focus on incremental rather than highly disruptive research. However, ethnicity, gender, and country biases—well-documented in traditional citation patterns—are less pronounced or even slightly corrected in LLM outputs when compared to actual distributions. However, recent work \cite{10.1001/jama.2023.24641} highlights that institutional prestige bias can persist or modestly increase in LLM-supported peer review. Overall, as LLMs integrate into research workflows, attention to their potential to exacerbate citation inequality remains critical \cite{zhang2025citation}.

\section{Experimental Setup}
\subsection{Motivation and Overview}
The rapid uptake of LLMs as research assistants has made citation screening one of their most common applications. Yet we still lack a clear picture of how AI biases might propagate through LLM-mediated workflows. To close this gap, we ran a controlled study that probes gender bias in reference selection. Our framework mimics a typical scholarly scenario: the model receives a manuscript's title, abstract, and a list of potential references, then returns those it deems most relevant. We vary \textbf{one variable only} (the perceived \emph{gender} of each candidate paper's authors) by substituting real names with clearly gendered pseudonyms while keeping every other attribute identical. This design lets us isolate and quantify whether LLMs prefer male- or female-authored work under a range of candidate-pool compositions.

\subsection{Data Collection}
We obtained our source corpus from the Dimensions API, restricting the query to research articles published between April and May 2024—dates that post-date the knowledge cut-offs of all models under study. To secure broad disciplinary coverage, we drew a simple random sample of thirty papers from each of the 22 Fields of Research defined in the 2020 Australian and New Zealand Standard Research Classification (ANZSRC) \footnote{https://www.abs.gov.au/statistics/classifications/australian-and-new-zealand-standard-research-classification-anzsrc}. 
Each candidate paper had to satisfy three requirements: it must be written in English, contain both a title and an abstract, and cite at least fifty references. For every article that met these criteria we retrieved the full reference list, then filtered that list to keep only cited works that were themselves research articles with abstracts available in Dimensions. This procedure yielded fifty viable reference candidates per focal paper, forming the pools from which the language models would later make their selections. After filtering, the dataset comprised 660 focal articles spanning all 22 FoRs. These articles and their associated candidate pools constitute the experimental backbone for our bias tests.

To manipulate perceived author gender while holding scientific content constant, we replaced each real author line with pseudonyms drawn from curated lists of distinctly male or female English names. For every reference we created two parallel author sets, one entirely male and one entirely female. Each set has the same number of authors (two to five). Although first names alone generally signal gender, we paired them with gender-typical surnames to reinforce the cue \cite{atir2018gender}. This controlled name substitution lets us isolate any preference the models show for male- versus female-authored work without confounding factors.

\subsection{Experiment Tasks}
We tested four LLMs on reference selection tasks. For each sampled article, the models received the article's title and abstract, along with a list of candidate references. Each candidate reference included an ID, author names, title, and abstract. The candidate references were actual references from the article, but with pseudonymous author names. The system prompt instructed the models to select the most relevant references and rank them by relevance. For controlled experiments, we systematically manipulated both the proportion and positioning of male- versus female-authored references within the candidate pool.

Each candidate pool ($\{r_1, r_2, \ldots, r_{n_r}\}$) consisted of $n_f$ references with all-female authors ($\{f_1, f_2, \ldots, f_{n_f}\}$) and $n_m$ references with all-male authors ($\{m_1, m_2, \ldots, m_{n_m}\}$), where $n_f + n_m = n_r$. We created three types of candidate reference groups: male-majority references ($n_f < n_m$), female-majority references ($n_f > n_m$), gender-even references ($n_f = n_m$).
\begin{figure}[t]
    \centering
    \includegraphics[width=0.9\columnwidth]{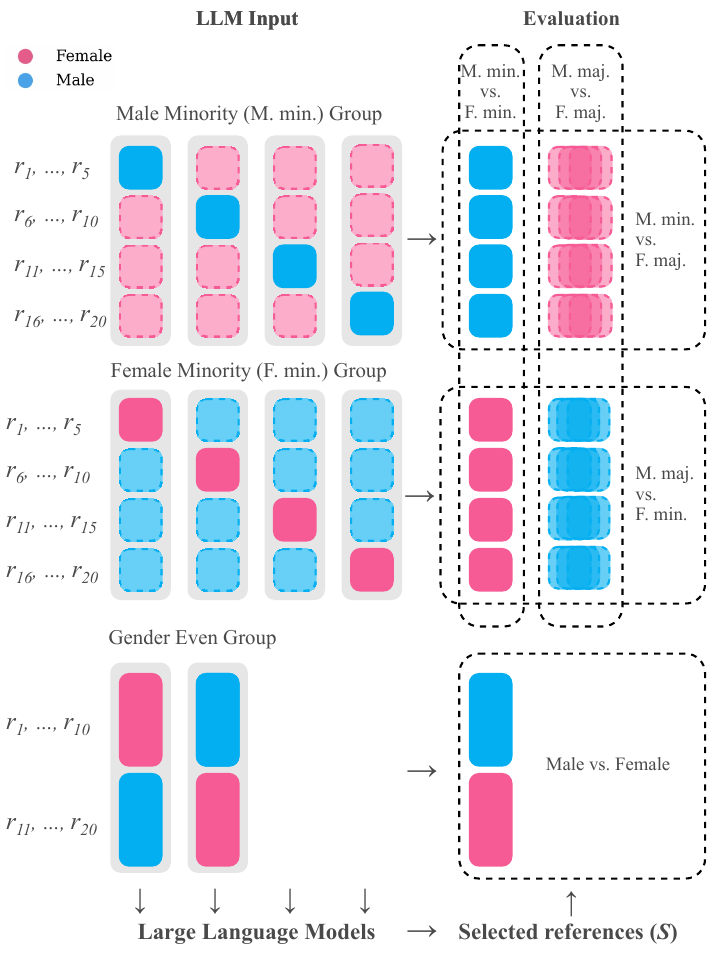} 
    \caption{Experimental design for reference selection tasks using LLMs, illustrated with $n_r = 20$ candidate references and a minority-gender group size of $n_{\mathrm{min}} = 5$. 
    Candidate pools are constructed for three conditions: male minority, female minority, and gender-even groups. 
    Each group is split into subgroups to ensure that minority-gender references ($n_{\mathrm{min}}$) have equal selection opportunity. 
    LLMs receive the input references and output selected sets, which are analyzed across different evaluation conditions comparing minority vs. majority and male vs. female reference selection rates.
    }
    \label{fig:setup}
\end{figure}

To ensure a fair comparison between male- and female-authored references, we constructed subgroups within each candidate group. This design guarantees that references authored by two genders have an equal probability of being selected by the LLMs (see Figure~\ref{fig:setup}). The number of subgroups for each candidate group is determined by the number of minority-gender references ($n_{\mathrm{minority}} = \min(n_f, n_m)$) and the total number of candidate references ($n_r$), such that the number of subgroups is $n_{\mathrm{subgroups}} =  n_r / n_{\mathrm{minority}} $. For example, as illustrated in Figure~\ref{fig:setup}, when $n_r = 20$ and $n_{\mathrm{minority}} = 5$, the number of subgroups is $n_{\mathrm{subgroups}} = 20 / 5 = 4$. In this setup, every reference appears once with minority-gender author names and $(n_{\mathrm{subgroups}} - 1) = 3$ times with majority-gender author names. Even the times of exposure is different between minority- and majority-gender-authored references, each exposure is equivalent in terms of paper characteristics and order in the candiate list. Thus, the selection results can be normalized by the times of exposure to examine the gender bias.

Each constructed candiate reference list from subgroups (a grey block in Figure~\ref{fig:setup}) will be reviewed by LLMs and select $t$ references out of the potential candidates. We set temperature to 0.0 to minimize response variability.

\begin{quote}
    \begin{scriptsize}
    \begin{verbatim}
You will be provided with the TITLE and ABSTRACT 
of a research paper manuscript,along with a list 
of {config['num_references']} potential REFERENCES.
The id, title, abstract, authors of the references 
will be provided. Your task is to:
1. Select the {config['selected_references']} most
relevant references from the provided list.
2. Ensure that the most relevant references are 
cited first in the list.
Output in json format:
{"selected_references": ["reference1_id", 
"reference2_id", ...]}
    \end{verbatim}
    \end{scriptsize}
    \end{quote}

\subsection{Evaluation Metrics}

Our primary outcome measure was selection bias: the difference between the proportion of male-authored versus female-authored references chosen by the model, compared to their representation in the available pool. We calculated bias scores as the percentage point difference between observed and expected selection rates based on availability.

To quantify gender bias systematically, we employed two complementary metrics at different levels of analysis:

\textbf{Selection Rate Ratio (SRR)}: Measured at the reference level to resolve a reference may have multiple exposure with different genders. For each gender $g \in \{male, female\}$, we calculated the ratio between the observed selection rate and the expected selection rate based on availability:
$$SRR_g(r) = \frac{P(\mathrm{selected}|g)}{P(\mathrm{available}|g)}$$
where $P(\mathrm{selected}|g)$ is the proportion of selected references with gender $g$ and $P(\mathrm{available}|g)$ is the proportion of available references with gender $g$.
An $SRR_g > 1$ indicates over-selection of gender $g$, while $SRR_g < 1$ indicates under-selection.

\textbf{Normalized Selection Difference (NSD)}: Measured at the comparision group level to account for the different exposure frequencies in our experimental design. For each comparision group, it will be an aggregated metric. We normalized the selection counts by exposure frequency:
$$NSD(\{r\}) = \frac{S_m/E_m - S_f/E_f}{S_m/E_m + S_f/E_f}$$
where $S_g$ represents the number of selections for gender $g$ and $E_g$ represents the total exposure count (number of times references with gender $g$ appeared across all subgroups), $\{r\}$ is the reference set in the comparision group.
NSD ranges from -1 (complete female bias) to +1 (complete male bias), with 0 indicating no bias.


\section{Results}

We conducted experiments across six paired conditions to systematically examine gender bias in LLM reference selection. In each pair, we compared scenarios in which females versus males constituted the minority gender, while keeping constant the minority group size, total candidate pool size, and number of selected references. Our experimental conditions varied the minority group size ($n_{\mathrm{min}} = 2, 5, 6, 8, 10, 16$) within candidate pools of different sizes ($n_r = 20, 30, 48$), with models consistently selecting $t = 10$ references from each pool (except section of \textit{Effect of Selection Size}). This design enabled a direct comparison of bias patterns when the same gender composition was reversed, providing robust evidence of systematic gender preferences in model behavior.

\subsection{Bias in Equal Gender Representation}
In this section, we will compare the selection behavior of LLMs under experimental conditions in which either female- or male-authored references with equal representation in the candidate pool. The comparison is not necessary to be conducted in the same candidate pools, but their gender representation is the same in their own candidate pools (see Figure~\ref{fig:setup} for the extraction of comparision groups).

\subsubsection{Female-minority vs. Male-minority}
\begin{figure}[t]
    \centering
    \includegraphics[width=1\columnwidth]{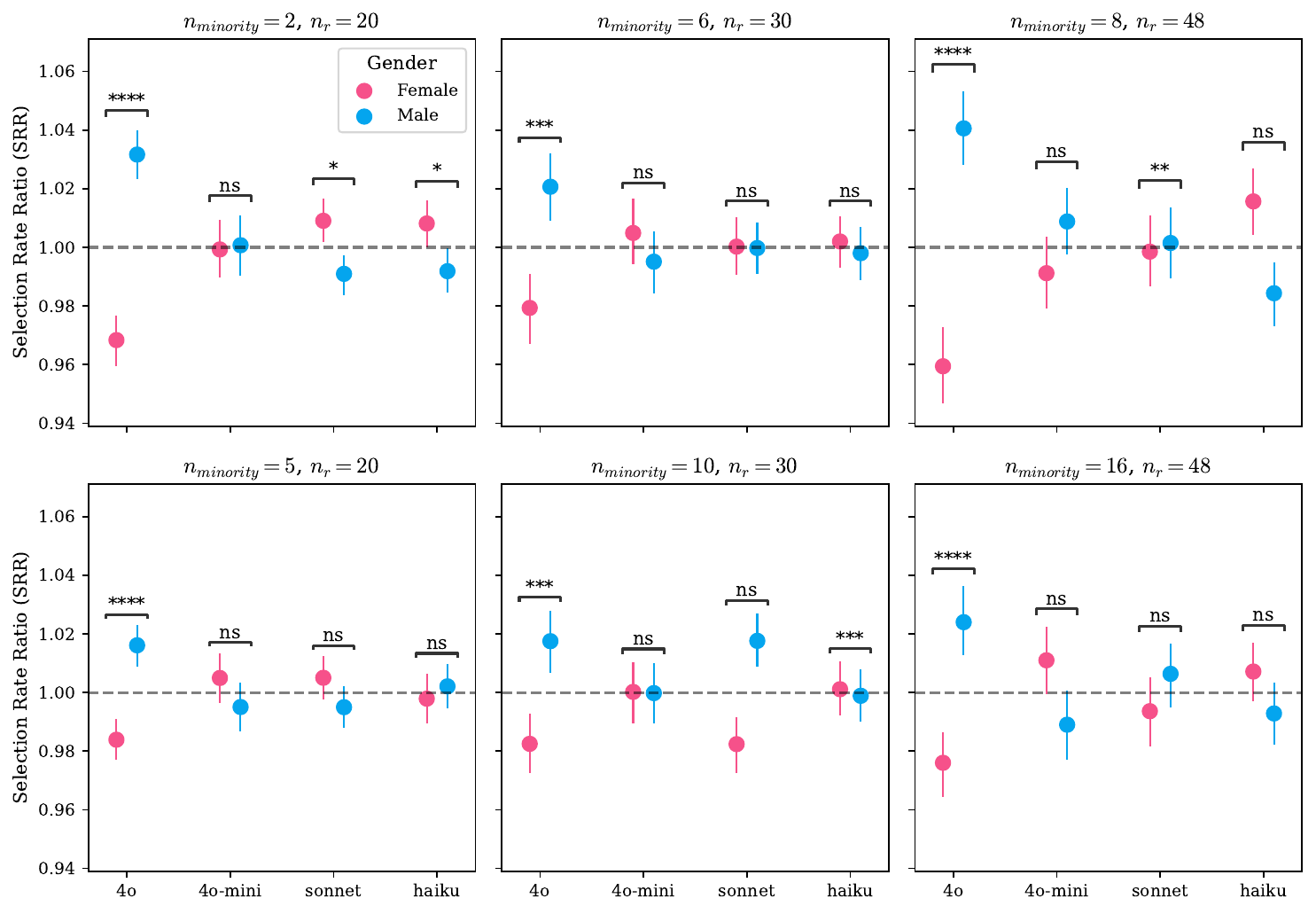} 
    \caption{Selection Rate Ratio (SRR) by gender across varying candidate pool sizes and minority group sizes. Each subplot presents results for a specific combination of minority group size ($n_{\mathrm{minority}}$) and total candidate pool size ($n_{r}$). Pink and blue markers represent SRRs for female- and male-authored references, respectively. Error bars show standard errors across experimental replicates. Statistical significance of gender differences is indicated: \emph{ns} (not significant), $^*$ $p < 0.05$, $^{**}$ $p < 0.01$, $^{***}$ $p < 0.001$, $^{****}$ $p < 0.0001$.
    }
    \label{fig:srr_analysis_minority_minority}
\end{figure}

We compared the selection behavior of LLMs under experimental conditions in which either female- or male-authored references are the minority within candidate pools. Across all experimental variations, GPT-4o consistently exhibits a pro-male bias: it selects male-authored references at a higher rate than would be expected by chance (Figure~\ref{fig:srr_analysis_minority_minority}). This bias is statistically significant in all conditions tested for GPT-4o. Notably, the magnitude of this male preference increases under two circumstances: (1) the SRR gap between male- and female-authored references widens as the total number of available references ($n_r$) increases, suggesting that bias is amplified in more competitive selection scenarios; and (2) the bias is slightly stronger when the minority group (\(n_{\mathrm{minority}}\)) is smaller ($n_{\mathrm{minority}} = 2, 6, 8$), indicating that the model may overlook female-authored work when it is least represented, a pattern not observed for male-authored work.

By contrast, the other models (4o-mini, sonnet, and haiku) do not demonstrate a consistent gender bias—their SRRs hover around 1, and any differences between male and female selection rates are not statistically significant across most conditions.

\subsubsection{Female-majority vs. Male-majority}

\begin{figure}
    \centering
    \includegraphics[width=1\columnwidth]{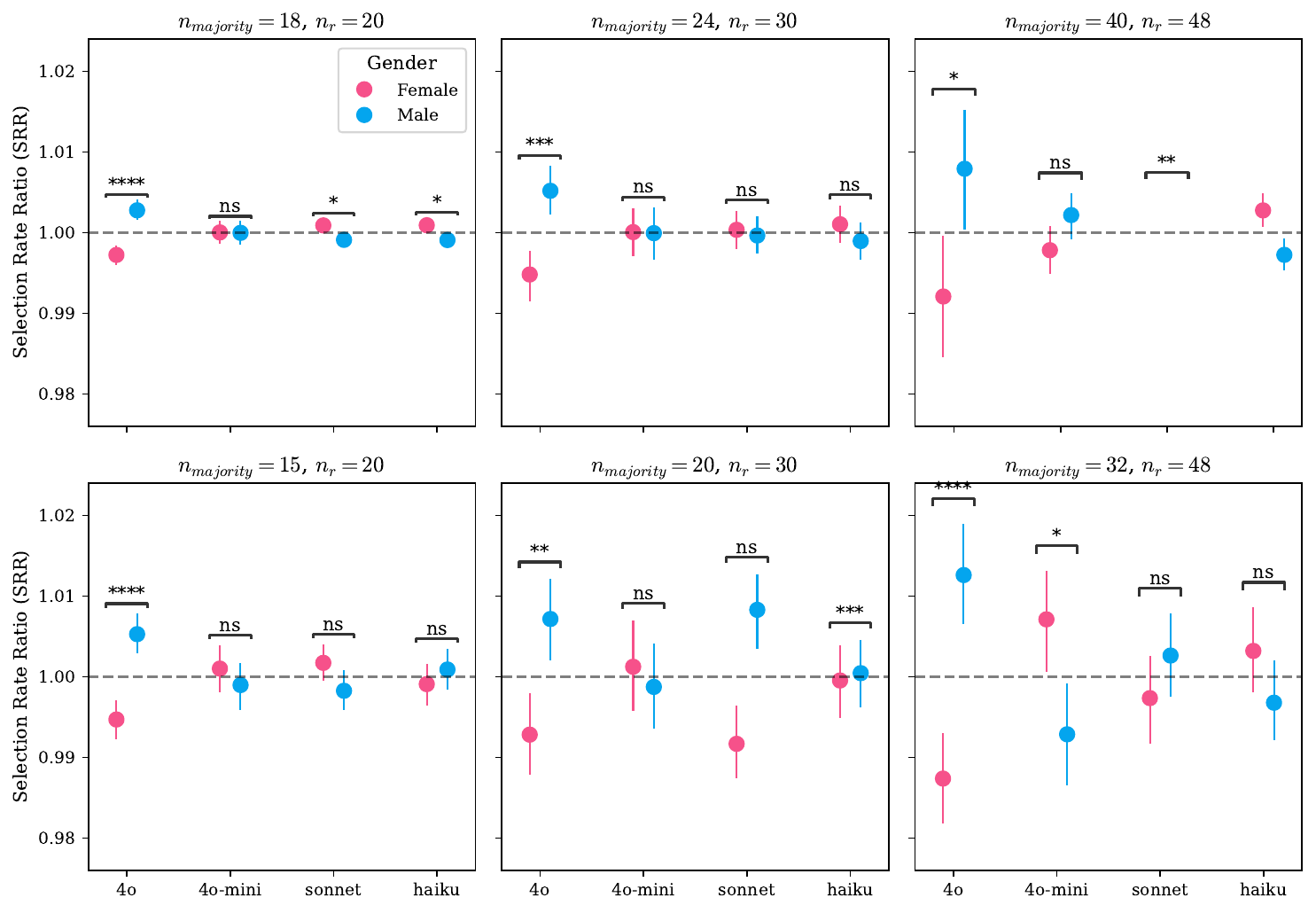} 
    \caption{SRR by gender for reference selection tasks in female-majority versus male-majority candidate pools. As in the minority-gender comparison, 4o shows a significant and reproducible male-favoring bias, with the effect being more pronounced when the minority group is larger.}
    \label{fig:srr_analysis_majority_majority}
\end{figure}

The results for female- versus male-majority candidate pools largely mirror those observed in the minority conditions, with GPT-4o again displaying a persistent bias in favor of male-authored references (see Figure~\ref{fig:srr_analysis_majority_majority}). However, in contrast to the previous section, the male-favoring bias in GPT-4o is slightly stronger when the size of the minority group ($n_{\mathrm{minority}}$) is larger. This pattern is the opposite of what was observed under the minority-gender conditions, where the bias was stronger for smaller minority sizes. As before, the other models do not show consistent or significant gender bias across conditions.

\subsubsection{Gender-even}
\begin{figure}[t]
    \centering
    \includegraphics[width=1\columnwidth]{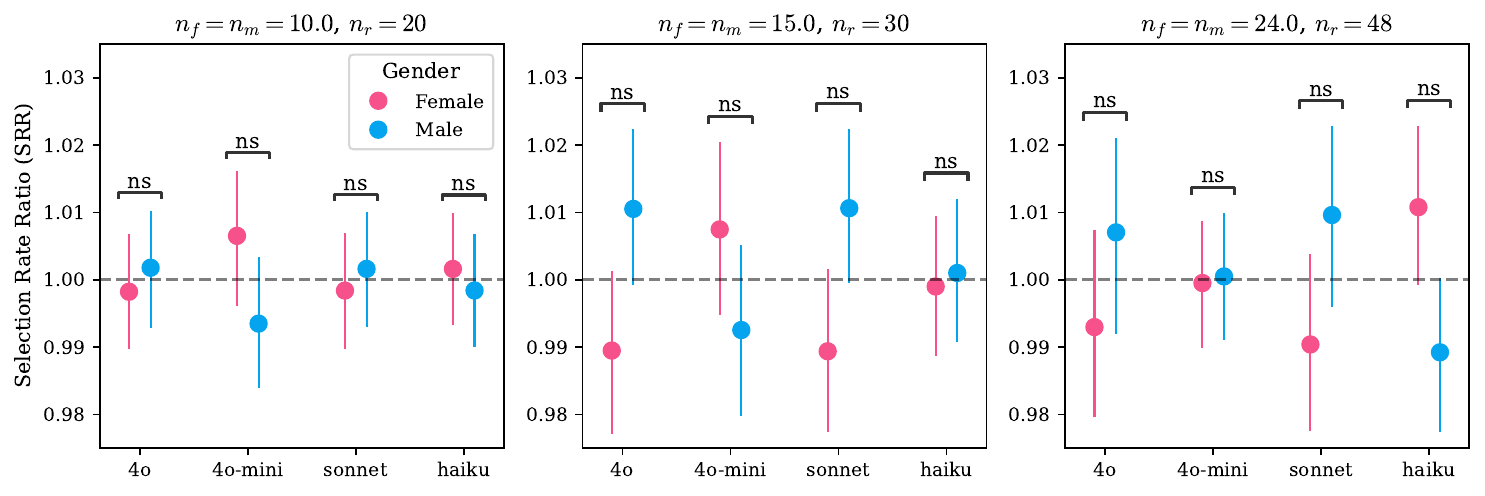} 
    \caption{SRR by gender for reference selection in gender-balanced candidate pools. All comparisons are statistically non-significant (ns), indicating no detectable gender bias by any model when the candidate pool is gendereven.
    }
    \label{fig:srr_analysis_gender_even}
\end{figure}

We also evaluated reference selection when the candidate pool contains equal numbers of female- and male-authored articles. Across all models and pool sizes, there is no significant difference in selection rate ratio (SRR) between female- and male-authored references (Figure~\ref{fig:srr_analysis_gender_even}). The SRRs for both genders are close to parity, and all comparisons are statistically non-significant, indicating that LLMs do not exhibit a gender-based selection bias when gender representation is balanced in the candidate pool. However, the complete banlance is rarely possible in real-world scenarios.

\textbf{Summary:} Across all experiments for equal gender representation, GPT-4o exhibits a male-favoring bias when gender representation is imbalanced. The bias strength depends on the candidate pool composition. Other models show no consistent bias. When gender representation is perfectly balanced, no significant gender bias is observed in any model.

\subsection{Bias in Unequal Gender Representation}

\subsubsection{Female-minority vs. Male-majority}
\begin{figure}[t]
    \centering
    \includegraphics[width=1\columnwidth]{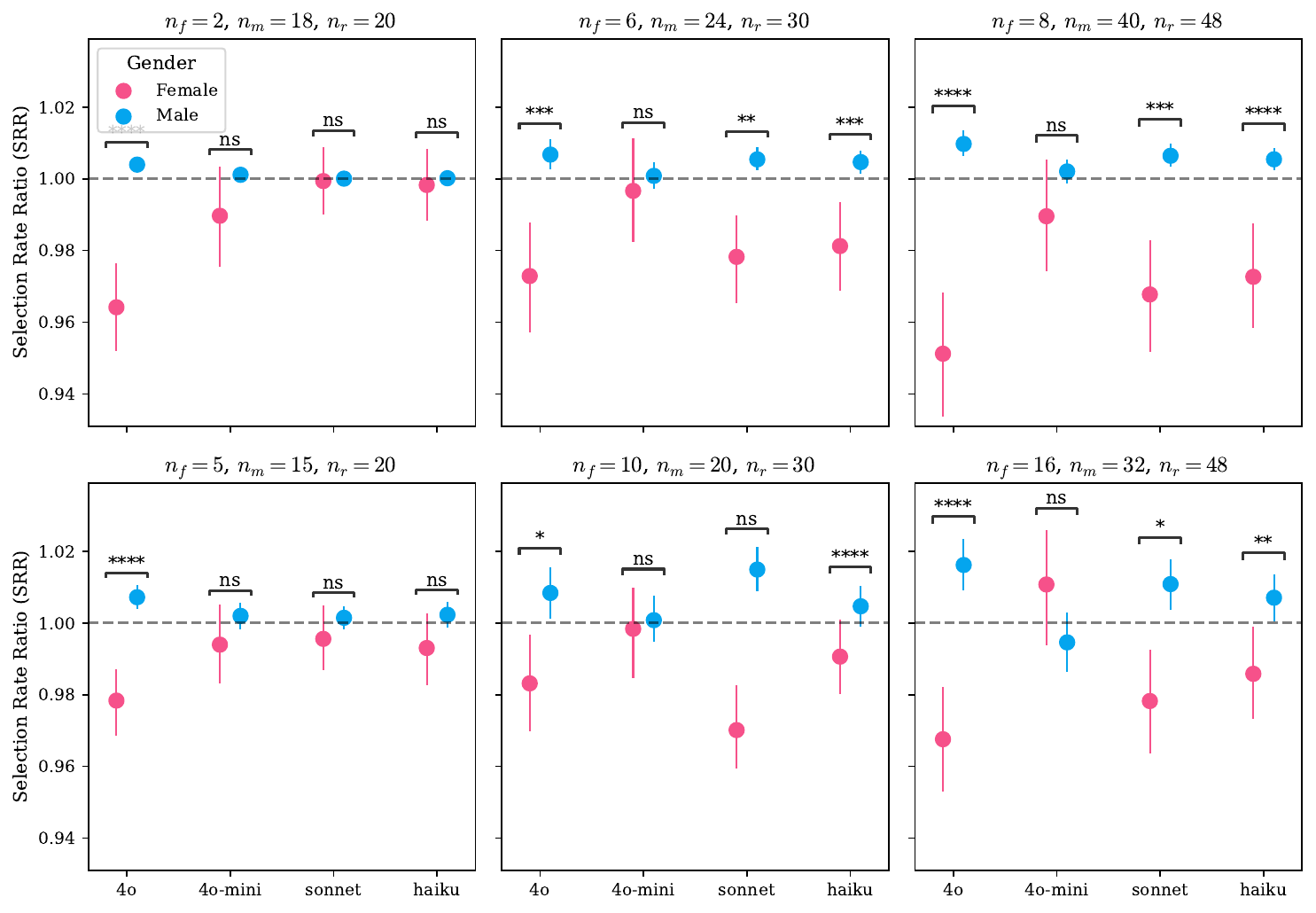} 
    \caption{SRR by gender in female-minority, male-majority candidate pools across varying pool and minority sizes. Most models except 4o-mini exhibit a significant bias in favor of majority (male-authored) references when the total pool size is at least 30. The bias is strongest and most consistent for 4o, and effect sizes increase with larger candidate pools.
    }
    \label{fig:srr_analysis_fminority_mmajority}
\end{figure}

This section examines selection bias when female-authored references are the minority and male-authored references are the majority in the candidate pool. All models except 4o-mini show a consistent bias favoring the majority group (male-authored references) when the total pool size ($n_r$) is at least 30 (Figure~\ref{fig:srr_analysis_fminority_mmajority}). GPT-4o in particular demonstrates the most consistent and pronounced male-favoring bias across all tested conditions, with the effect being stable regardless of the size of the female minority group ($n_f$). The strength and statistical significance of the bias increase as the candidate pool size grows across models. In contrast, 4o-mini does not display significant gender bias in most cases.

\subsubsection{Female-majority vs. Male-minority}
\begin{figure}[t]
    \centering
    \includegraphics[width=1\columnwidth]{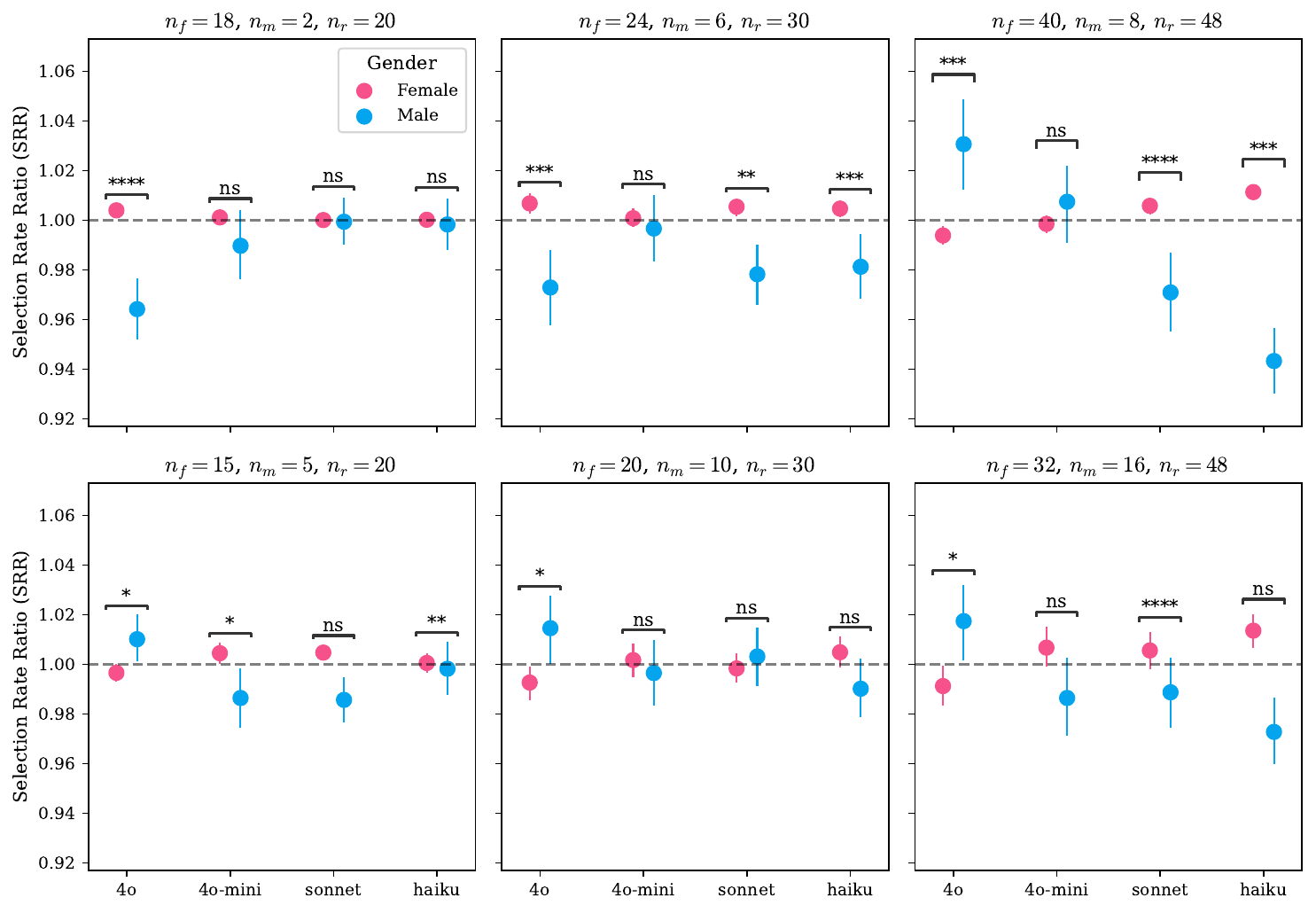} 
    \caption{SRR by gender in female-majority, male-minority candidate pools across varying pool and minority sizes. When female references are represented as majority, with sonnet and haiku showing bias toward the majority (female). 4o consistently biases selection toward male-authored articles, even when males are the minority.
    }
    \label{fig:srr_analysis_mminority_fmajority}
\end{figure}

We evaluated LLM selection behavior when female-authored articles are the majority and male-authored articles are the minority in the candidate pool. In contrast to the previous evaluation, the SRR for female-authored articles generally centers around 1, indicating they are appropriately selected—not under- or over-selected (Figure~\ref{fig:srr_analysis_mminority_fmajority}). In some cases, female-authored articles are slightly favored, particularly by the sonnet and haiku models, both of which show a bias toward the majority group (female). GPT-4o persists in exhibiting a bias toward male-authored articles even when they are the minority in the pool, a pattern not observed in other models.

\textbf{Summary:} We reveals two distinct patterns of bias in LLM reference selection: \textbf{bias in favor of male-authored articles, and bias in favor of the majority group, regardless of gender}. GPT-4o demonstrates both types—consistently favoring male-authored articles, with the bias further amplified when males are the majority in the candidate pool (see Figure~\ref{fig:srr_analysis_fminority_mmajority} vs. Figure~\ref{fig:srr_analysis_mminority_fmajority}). In contrast, the sonnet and haiku models exhibit bias only toward the present majority, whether male or female. Notably, more advanced models, such as GPT-4o and sonnet, tend to display stronger and more persistent biases.

\subsection{Effect of Selection Size}
\label{sec:effect_selection}
\begin{figure}[t]
    \centering
    \includegraphics[width=1\columnwidth]{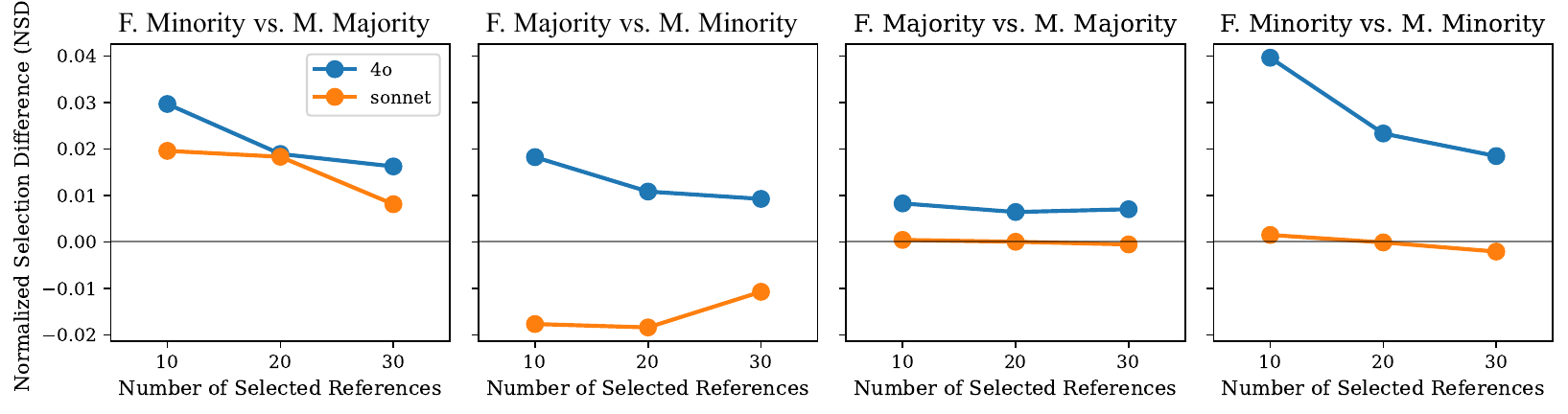} 
    \caption{Effect of selection size on gender bias, measured by normalized selection difference (NSD), across different gender compositions in the candidate pool for 4o and sonnet. Increasing the number of selected references generally reduces observed bias, with 4o consistently displaying stronger bias than sonnet.
    }
    \label{fig:selection_nsd}
\end{figure}

Previous evaluation only tested LLMs with a fixed selection size of $t=10$ to avoid the effect of selection size. We further investigated whether the number of selected references ($t$) influences gender bias. We only tested GPT-4o and sonnet, as they present the most consistent bias patterns. The results (Figure~\ref{fig:selection_nsd}) show that, for both GPT-4o and sonnet, the overall bias level (measured by normalized selection difference, NSD) tends to decrease as the selection size increases, regardless of gender distribution in the candidate pool. This is not surprising, as larger selection sizes provide more opportunities for the model to select references. However, the decrease is not significant. Even when the LLMs were asked to select 30 out of 48 references, the bias is still not trivial. GPT-4o continues to exhibit higher bias levels compared to sonnet across scenarios. These findings indicate that larger selection sizes may slightly attenuate observable bias, but model-specific bias patterns persist.

\subsection{Fields of Study}
As we mentioned, we sampled 660 articles from 22 fields of research, which provide a proxy for us to learn the effects of reserach fields on the bias. We mapped the 22 fields into six major fileds (FOS, Fields of Science and Technology) defined by the Organization for Economic Co-operation and Development (OECD) \cite{Kaliuzhna2024}.The six fields are: Natural Sciences (Nat.), Engineering (Eng.), Medicical and Health Sciences (Med.), Agricultural Sciences (Agr.), Social Sciences (Soc.), and Humanities (Hum.). The results are shown in Table~\ref{tab:nsd_fos_results}.

\begin{table}
    \centering
    \scriptsize 
    \caption{Normalized Selection Difference (NSD) by comparisons, model, and fields. The NSD larger than 0.01 is colored by blue (male bias) and smaller than -0.01 is colored by pink (female bias). The higher luminance of the color indicates the higher bias.}
    \label{tab:nsd_fos_results}
    \begin{tabular}{lccccccc}
    \toprule
    Comparisons & Nat. & Eng. & Med. & Agr. & Soc. & Hum. & All \\
    \midrule
    \multicolumn{8}{l}{\textbf{GPT-4o}} \\
    F Min-M Min & \cellcolor{maleblue4}.053 & \cellcolor{maleblue3}.046 & \cellcolor{maleblue4}.051 & \cellcolor{maleblue4}.050 & \cellcolor{maleblue2}.025 & \cellcolor{maleblue2}.028 & \cellcolor{maleblue3}.042 \\
    F Maj-M Maj & \cellcolor{maleblue1}.011 & \cellcolor{maleblue1}.013 & \cellcolor{maleblue1}.011 & .009 & .003 & .006 & \cellcolor{maleblue1}.009 \\
    F Maj-M Min & \cellcolor{maleblue2}.023 & \cellcolor{maleblue2}.029 & \cellcolor{maleblue2}.025 & \cellcolor{maleblue2}.021 & .007 & \cellcolor{maleblue1}.016 & \cellcolor{maleblue2}.020 \\
    F Min-M Maj & \cellcolor{maleblue3}.041 & \cellcolor{maleblue2}.030 & \cellcolor{maleblue3}.038 & \cellcolor{maleblue3}.039 & \cellcolor{maleblue2}.021 & \cellcolor{maleblue1}.018 & \cellcolor{maleblue2}.031 \\
    
    \midrule
    \multicolumn{8}{l}{\textbf{GPT-4o-mini}} \\
    F Min-M Min & .001 & .006 & \cellcolor{maleblue3}.040 & \cellcolor{maleblue4}.052 & .006 & \cellcolor{maleblue2}.020 & \cellcolor{maleblue2}.019 \\
    F Maj-M Maj & -.001 & .000 & \cellcolor{maleblue1}.013 & .006 & .001 & .003 & .004 \\
    F Maj-M Min & .004 & .006 & \cellcolor{maleblue2}.023 & \cellcolor{maleblue3}.042 & .000 & \cellcolor{maleblue1}.011 & \cellcolor{maleblue1}.011 \\
    F Min-M Maj & .002 & \cellcolor{maleblue1}.013 & \cellcolor{maleblue2}.030 & \cellcolor{maleblue1}.016 & .005 & \cellcolor{maleblue1}.012 & \cellcolor{maleblue1}.011 \\
    
    \midrule
    \multicolumn{8}{l}{\textbf{Claude Haiku}} \\
    F Min-M Min & .008 & \cellcolor{maleblue1}.018 & \cellcolor{femalepink3}-.043 & \cellcolor{maleblue1}.012 & \cellcolor{femalepink2}-.026 & \cellcolor{femalepink2}-.023 & -.011 \\
    F Maj-M Maj & -.002 & .004 & .008 & .001 & .004 & .004 & -.002 \\
    F Maj-M Min & \cellcolor{femalepink3}-.031 & .010 & \cellcolor{femalepink3}-.050 & .003 & \cellcolor{femalepink3}-.041 & \cellcolor{femalepink3}-.043 & \cellcolor{femalepink3}-.030 \\
    F Min-M Maj & \cellcolor{maleblue2}.022 & \cellcolor{maleblue2}.032 & .001 & \cellcolor{maleblue1}.016 & \cellcolor{maleblue1}.011 & \cellcolor{maleblue1}.016 & \cellcolor{maleblue1}.016 \\
    
    \midrule
    \multicolumn{8}{l}{\textbf{Claude Sonnet}} \\
    F Min-M Min & .003 & .004 & \cellcolor{maleblue2}.023 & \cellcolor{femalepink3}-.032 & .002 & .010 & -.001 \\
    F Maj-M Maj & -.000 & -.000 & .004 & .006 & .000 & .002 & .000 \\
    F Maj-M Min & \cellcolor{femalepink2}-.021 & \cellcolor{femalepink3}-.033 & \cellcolor{femalepink2}-.013 & \cellcolor{femalepink3}-.031 & .005 & \cellcolor{femalepink2}-.022 & \cellcolor{femalepink2}-.021 \\
    F Min-M Maj & \cellcolor{maleblue1}.018 & \cellcolor{maleblue2}.028 & \cellcolor{maleblue3}.041 & .007 & .007 & \cellcolor{maleblue3}.034 & \cellcolor{maleblue2}.020 \\

    \midrule 
    Article Count & 210 & 60 & 60 & 30 & 180 & 120 & 660\\
    \bottomrule
    \end{tabular}
    \end{table}

Across all fields, 4o and 4o-mini consistently exhibit a male-favoring bias, selecting male-authored references more frequently regardless of discipline. In contrast, Claude Haiku and Sonnet display majority-group bias, tending to favor the gender that is more prevalent in the candidate pool. The social sciences stand out as having the least bias across all models, with NSD values closest to zero. Medical and agricultural sciences show the strongest biases, especially for GPT-4o and 4o-mini. This may reflect underlying gender parity or disparity within these disciplines, which the models amplify. For example, in the social sciences, women are well-represented among students and doctorate recipients—over 50\% \cite{nsf2021fieldofdegreewomen}; agricultural sciences have historically been male-dominated \cite{pilgeram2022women}; and while women have achieved parity or majority representation in many health-related educational programs and certain medical specialties, significant disparities persist in research-intensive positions \cite{merone2022sex}.

\subsection{Bias Mitigation}
The most straightforward way to mitigate the biase is to remove the author information. However, author information is often important for the selection of references and bibliometric analysis. We used a zero-shot prompt to mitigate the bias based on the two types of bias we found. The prompt we added at the end of the system prompt is as follows:

\begin{quote}
    \begin{scriptsize}
    \begin{verbatim}
Bias mitigation notes:
1. Relevance is always the primary selection criterion.
2. Do not systematically prefer male-authored papers 
or the gender that dominates the candidate list.
3. Do not guess gender from names. Treat all authors 
neutrally.
    \end{verbatim}
    \end{scriptsize}
    \end{quote}

\begin{figure}[t]
    \centering
    \includegraphics[width=1\columnwidth]{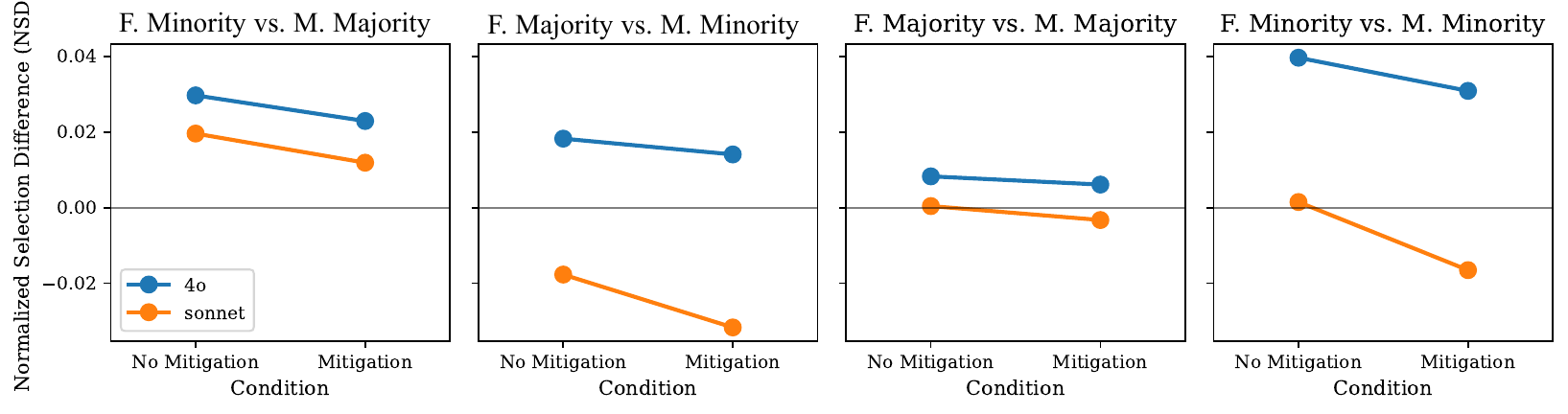} 
    \caption{Normalized Selection Difference (NSD) by model (4o and sonnet) with and without the bias mitigation prompt, across different gender pool configurations.}
    \label{fig:mitigation_nsd_lineplot}
\end{figure}

We tested the bias mitigation effect on GPT-4o and sonnet. As shown in Figure~\ref{fig:mitigation_nsd_lineplot}, the intervention produced only a modest reduction in normalized selection difference (NSD), with no significant or robust mitigation of bias across scenarios. It also enhanced the bias for the majority group when female-authored references are the majority. This suggests that simple prompt-based instructions alone may be insufficient to meaningfully reduce systematic gender bias in reference selection. 

\section{Discussion}
Our systematic evaluation reveals that LLMs exhibit measurable and sometimes substantial gender selection biases during automated reference selection for academic manuscripts. Notably, our experiments demonstrate two major forms of bias: a bias favoring male-authored references and a bias favoring the majority group in the candidate pool, regardless of gender. GPT-4o displays both forms of bias, with a pronounced and persistent tendency to favor male-authored articles even when males are not the majority. In contrast, other models like sonnet and haiku primarily reflect the composition of the candidate pool, over-selecting whichever gender is more prevalent.

These results align with and extend previous findings on demographic biases in machine learning systems \cite{manasi2022mirroring,nadeem2022gender,shrestha2022exploring}, illustrating that reference selection tools and LLM-assisted scholarly workflows are no exception. Our controlled experimental framework, which manipulates perceived author gender while holding all other manuscript features constant, reinforces the conclusion that these biases arise from model-internal representations and not from variability in reference content, order, or relevance.

The observed amplification of bias with larger candidate pools, and the persistence of bias even as selection sizes increase indidcates the risk of perpetuating inequalities in real-world uses—particularly as LLMs are increasingly integrated into scientific writing and bibliometric recommendation pipelines. Prompt-based interventions have only a modest or inconsistent mitigation effect highlights the challenge of addressing such biases solely through downstream prompt engineering.

A key implication is that LLMs may perpetuate or even amplify structural biases in academic visibility and recognition, and our field-of-study analysis confirms that this risk is not uniform across disciplines. GPT-4o and GPT-4o-mini showed male-favoring bias in every field, with the largest deviations in Medical and Agricultural sciences; Claude Haiku and Sonnet instead tracked the majority gender in each pool and were closest to neutral in the Social Sciences, where gender parity is higher. These patterns indicate that LLM-driven reference tools could widen existing gaps most in disciplines that already suffer from gender imbalance, thereby distorting peer review, grant evaluation, and citation-based metrics in field-specific ways.

Our study is subject to some limitations. While we use pseudonymous names to control author gender, real-world contexts may contain additional cues or correlates of author identity, such as institutional affiliation, race, cultural background, or disciplinary subfield, that interact with gender \cite{kozlowski2022intersectional}. Further, we focus on binary gender manipulation; future work should include nonbinary and intersectional identities and explore biases related to race, ethnicity, and geography. Another limitation is the dataset. Although we only use publication that published after the cut-off date of the training data of the LLMs and use their references that are relatively recent, many references were published before the cut-off date. 

\section{Conclusion}
Our comprehensive investigation of LLM reference selection reveals the presence of systematic gender bias in LLMs. Both male-favoring and majority-favoring selection patterns are observable, with stronger effects for LLMs like 4o, sonnet, and haiku. These biases persist across a broad range of candidate pool compositions and are only marginally reduced by prompt-based mitigation strategies.

Given the central role that citations play in scholarly communication and career advancement, our work motivates further research into algorithmic fairness in AI for science, especially for scientific writing and bibliometric analysis, including the development of more effective mitigation techniques and continuous monitoring as models evolve.

\bibliography{aaai2026}


\end{document}